\documentclass[12pt,a4paper]{article}
\usepackage{amsfonts,latexsym}
\usepackage{graphicx,color}
\usepackage{dcolumn}
\usepackage{graphicx}
\usepackage{amsmath}
\usepackage{amsfonts}
\usepackage{amssymb}

\renewcommand{\thefootnote}{\fnsymbol{footnote}}

\begin{document}

\vspace{12mm}

\begin{center}
{{{\Large {\bf Stability of scalarized charged black holes in the Einstein-Maxwell-Scalar theory}}}}\\[10mm]

{Yun Soo Myung$^a$\footnote{e-mail address: ysmyung@inje.ac.kr} and De-Cheng Zou$^{a,b}$\footnote{e-mail address: dczou@yzu.edu.cn}}\\[8mm]

{${}^a$Institute of Basic Sciences and Department  of Computer Simulation, Inje University Gimhae 50834, Korea\\[0pt] }

{${}^b$Center for Gravitation and Cosmology and College of Physical Science and Technology, Yangzhou University, Yangzhou 225009, China\\[0pt]}
\end{center}
\vspace{2mm}

\begin{abstract}
We analyze  the stability of scalarized charged  black holes  in the Einstein-Maxwell-Scalar (EMS) theory with quadratic coupling.
These  black holes are labelled  by the number of $n=0,1,2,\cdots$, where $n=0$ is called the fundamental
black hole and $n=1,2,\cdots$ denote the $n$-excited black holes.
We show that   the $n=0$  black hole is stable against full perturbations, whereas the $n=1,2$ excited
black holes are unstable against  the $s(l=0)$-mode scalar perturbation.  This is consistent
with the EMS theory with exponential coupling, but it contrasts to the $n=0$ scalarized black hole in the Einstein-Gauss-Bonnet-Scalar theory with quadratic coupling.
This implies that the endpoint of unstable  Reissner-Nordstr\"{o}m black holes with $\alpha>8.019$
is the  $n=0$  black hole with the same $q$.  Furthermore, we  study the scalarized charged black holes  in the EMS theory with scalar mass
$m^2_\phi=\alpha/\beta$.
\end{abstract}
\vspace{5mm}

\vspace{1.5cm}

\hspace{11.5cm}
\newpage
\renewcommand{\thefootnote}{\arabic{footnote}}
\setcounter{footnote}{0}


\section{Introduction}
A scalarization of the Reissner-Nordstr\"{o}m (RN) black holes  was obtained   in the
Einstein-Maxwell-Scalar (EMS) theory~\cite{Herdeiro:2018wub}.
The EMS theory is a simple second-order theory providing   three kinds of propagating modes of scalar, vector, and tensor around the black hole background.
It is worth reminding  that the appearance of the scalarized charged  black holes is closely connected to the
instability of the RN black hole~\cite{Myung:2018vug}.
We note that these black holes are denoted as  the $n=0,1,2,\cdots$ black holes with $\alpha$ coupling constant.

All black hole solutions could be linearly tested  to confirm that some solutions are selected  as black holes in the curved spacetimes.
Concerning the stability of scalarized black holes, it was firstly shown that the $n=0$ black hole is stable against $l=0(s$-mode) scalar perturbation, while $n=1,2,\cdots$ black holes are unstable against the $s$-mode scalar perturbation  in the Einstein-Born-Infeld-Scalar  theory~\cite{Doneva:2010ke}.
As was mention in~\cite{Blazquez-Salcedo:2018jnn},  a difference between exponential and quadratic couplings in
the Einstein-Gauss-Bonnet-Scalar (EGBS) theory is   that the $n=0$  black hole is stable against radial perturbations for the exponential coupling, while it  is unstable for the quadratic coupling. This implies that the $n=0$ black hole could be regarded as  the endpoint of the evolution of unstable Schwarzschild black hole for the exponential coupling, whereas this is not the case for the quadratic coupling.  Recently,  it is argued that the quadratic term controls the onset of the instability giving the $n=0$  black hole, while the higher-order terms  including the exponential coupling  control the stability of the $n=0$ black hole in the EGBS theory~\cite{Silva:2018qhn}.   Very recently, the spontaneous scalarization of black holes and its stability in the EGBS theory were studied by including a massive scalar term  for different couplings~\cite{Macedo:2019sem,Doneva:2019vuh}.

For the stability of scalarized black holes in the EMS theory with exponential coupling~\cite{Myung:2018jvi}, it is known that  the $n=0$ black hole is stable against full perturbations, while $n=1,2$ black holes are unstable against the $s$-mode scalar perturbation. In this case, the endpoint of unstable RN black holes may be  the stable $n=0$ black hole with the same $q$ in the EMS theory with exponential coupling. Hence, it is curious to know the stability issue of the $n=0,1,2$ black holes
in the EMS theory with quadratic coupling. In this respect, it is shown that the $n=0$ black hole may be stable in the EMS theory with quadratic coupling by mentioning   the positive potentials~\cite{Fernandes:2019rez}.

 In this work,
we will  study  the $n=0,1,2$ scalarized  charged black holes in the EMS theory with quadratic coupling
by observing the potentials and computing quasinormal mode spectrum. Also,  we  wish to investigate the scalarized charged black holes  in the EMS theory with scalar mass
$m^2_\phi=\alpha/\beta$.
The full tensor-vector-scalar perturbations will be adopted for the massless case.
Observing the potentials around the $n=0,1,2$ black holes and together with computing quasinormal frequencies of the five  physical modes,
we show that the $n=0$ black hole is still stable against full perturbations, while $n=1,2$ black holes
are unstable against the $s$-mode scalar perturbation in the EMS theory with quadratic coupling. This implies that the endpoint of unstable RN black holes with $\alpha>8.019$
and $q=Q/M=0.7$ may be  the  $n=0(\alpha \ge 8.019)$ scalarized charged  black hole with the same $q$.

\section{$n=0,1,2,\cdots$ black holes} \label{sec1}

We consider the action of EMS theory with quadratic coupling~\cite{Herdeiro:2018wub}
\begin{equation}
S_{\rm EMS}=\frac{1}{16 \pi}\int d^4 x\sqrt{-g}\Big[ R-2\partial_\mu \phi \partial^\mu \phi-V_\phi-(1+\alpha\phi^2) F^2\Big],\label{Action1}
\end{equation}
where $\alpha$ is a Maxwell-scalar coupling constant and we choose $V_\phi=0$. If one considers  a quadratic coupling of $\alpha \phi^2$,
one has to choose $\bar{\phi}={\rm const}$ to obtain the RN black hole with a different charge $\tilde{Q}^2= \bar{\phi}^2 Q^2$.
In order to make the analysis clear, here, we choose an equivalent coupling of  $1+\alpha \phi^2$~\cite{Fernandes:2019rez}
together with $\bar{\phi}=0$ to give the same RN black hole.

From the action (\ref{Action1}), the equations of motion are obtained as
\begin{eqnarray}
 G_{\mu\nu}=2\partial _\mu \phi\partial _\nu \phi -(\partial \phi)^2g_{\mu\nu}+2T_{\mu\nu} \label{equa1}
\end{eqnarray}
with $G_{\mu\nu}=R_{\mu\nu}-(R/2)g_{\mu\nu}$ and  $T_{\mu\nu}=(1+\alpha\phi^2)(F_{\mu\rho}F_{\nu}~^\rho-F^2g_{\mu\nu}/4)$,
and the Maxwell equation takes the form
\begin{equation} \label{M-eq}
\nabla^\mu F_{\mu\nu}-2\alpha \phi\nabla^{\mu} (\phi)F_{\mu\nu}=0.
\end{equation}
The scalar equation is given by
\begin{equation}
\square \phi -\frac{\alpha F^2}{2} \phi=0 \label{s-equa}.
\end{equation}

 We introduce  the  scalar perturbed equation [$(\bar{\square}+ \alpha Q^2/r^4) \delta \varphi= 0$] on the RN black hole background
\begin{equation} \label{RN-sol}
ds^2_{\rm RN}=-\tilde{N}(r)e^{-\tilde{\delta}(r)}dt^2+\frac{dr^2}{\tilde{N}(r)}+r^2(d\theta^2+\sin^2\theta d\varphi^2)
\end{equation}
with
\begin{equation}\label{RN-sols}
\tilde{N}(r)=1-\frac{2M}{r}+\frac{Q^2}{r^2},~ \tilde{\delta}(r)=0,~\tilde{\phi}(r)=0,~\tilde{A}_0=\frac{Q}{r}.
\end{equation}
We note that this RN background  is surely independent of $\alpha$.
Considering the separation of variables around the spherically symmetric RN background
\begin{equation} \label{mscalar-sp}
\delta \varphi(t,r,\theta,\varphi)=\frac{u(r)}{r}e^{-i\omega t}Y_{lm}(\theta,\varphi),
\end{equation}
and introducing a tortoise coordinate $r_*$ defined by $dr_*=dr/\tilde{N}(r)$, the perturbed scalar  equation is given by
\begin{equation} \label{msch-2}
\frac{d^2u}{dr_*^2}+\Big[\omega^2-V_{\rm ml}(r)\Big]u(r)=0,
\end{equation}
where the massless potential takes the form
\begin{equation} \label{mpot-c}
V_{\rm ml}(r)=\tilde{N}(r)\Big[\frac{2M}{r^3}+\frac{l(l+1)}{r^2}-\frac{2Q^2}{r^4}-\alpha\frac{Q^2}{r^4}\Big].
\end{equation}
Actually, (\ref{msch-2}) is suitable for analyzing the stability of RN black hole.

In order to obtain bifurcation points, one needs to solve the static perturbed equation for $\varphi(r)=u(r)/r$ as
\begin{equation} \label{ssp-eq}
\frac{1}{r^2}\frac{d}{dr}\Big[r^2 \tilde{N}(r)\frac{d\varphi(r)}{dr}\Big]-\Big[\frac{l(l+1)}{r^2}-\frac{\alpha Q^2}{r^4}\Big] \varphi(r)=0.
\end{equation}
Here, Eq.(\ref{ssp-eq}) describes an eigenvalue problem: for  given $l=0$, requiring an asymptotically vanishing, smooth scalar
selects a discrete set of the bifurcation points for scalarized solution
as $\alpha_n(q=0.7)=\{8.019,~40.84,~99.89,\cdots\}$.
In this case, the bifurcation points of the RN solution are the same as those of exponential
coupling $e^{\alpha \phi^2}$ \cite{Myung:2018vug,Myung:2018jvi}
because the static scalar perturbed   equation takes the same form as in (\ref{ssp-eq}).
In Fig. 1, these solutions are classified
by the node number $n$ for $\varphi(z)$ with $z=r/(2M)$. Furthermore, $n$ will
denote the order number for classifying different branches of scalarized black holes.

\begin{figure*}[t!]
  \centering
   \includegraphics{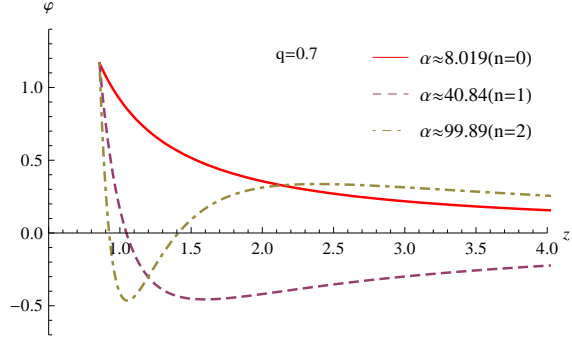}
\caption{Radial profiles of $\varphi(z)$ as function of $z=r/(2M)$ for the first three perturbed scalar solutions on the RN black hole with $q=0.7$.
Here $n$ represents the  node number of  $\varphi(z)$ and it will  denote the order number for labelling  different scalarized black holes.  }
\end{figure*}

To obtain scalarized charged black holes, we have to introduce  the spherically symmetric metric ansatz as
\begin{eqnarray}\label{nansatz}
ds^2_{\rm SBH}=\bar{g}_{\mu\nu}dx^\mu dx^\nu=-N(r)e^{-2\delta(r)}dt^2+\frac{dr^2}{N(r)}+r^2(d\theta^2+\sin^2\theta d\varphi^2)
\end{eqnarray}
with a metric function of $N(r)=1-2m(r)/r$, in addition to  electric potential  $\bar{A}_0=v(r)$ and scalar field $\bar{\phi}(r)$.
We note that  scalarized charged black holes could be obtained  by restricting an allowable range for $\alpha$.
The threshold of  instability for the RN black hole is closely related to the appearance
of the $n=0(\alpha\ge 8.019)$  black hole. Also, we emphasize that the static scalar perturbation around the RN black hole determines the appearance of $n=1,2\cdots$  black holes.

Plugging (\ref{nansatz}) into (\ref{equa1})-(\ref{s-equa}), one has  the four equations
\begin{eqnarray}
&&-2m'(r)+e^{2\delta(r)}\Big[1+\alpha(\bar{\phi}(r))^2\Big]r^2(v'(r))^2+[r^2-2rm(r)](\bar{\phi}'(r))^2=0,\label{neom1}\\
&&\delta'(r)+r(\bar{\phi}'(r))^2=0,\label{neom2}\\
&&v'(r)\Big[2+r\delta'(r)+\frac{2r\alpha\bar{\phi}(r)\bar{\phi}'(r)}{1+\alpha\bar{\phi}(r)^2}\Big]+r v''(r)=0,\label{neom3}\\
&&e^{2\delta(r)}r^2\alpha\bar{\phi}(r)(v'(r))^2+r[r-2m(r)]\bar{\phi}''(r)\nonumber\\
&&-\Big[m(r)(2-2r\delta'(r))
+r(-2+r\delta'(r)+2m'(r))\Big]\bar{\phi}'(r)=0, \label{neom4}
\end{eqnarray}
where the prime ($'$) denotes differentiation with respect to its argument.

Considering the existence of a horizon located at $r=r_+$,  one suggests  an
approximate solution to equations in the near horizon
\begin{eqnarray}
&&m(r)=\frac{r_+}{2}+m_1(r-r_+)+\ldots,\label{apsc-1}\\
&&\delta(r)=\delta_0+\delta_1(r-r_+)+\ldots,\label{aps-2}\\
&&\bar{\phi}(r)=\phi_0+\phi_1(r-r_+)+\ldots,\label{aps-3}\\
&&v(r)=v_1(r-r_+)+\ldots,\label{aps-4}
\end{eqnarray}
where the four coefficients are given by
\begin{eqnarray}\label{ncoef}
&&m_1=\frac{Q^2}{2r_+^2(1+\alpha\phi_0^2)},\quad
\phi_1=\frac{\alpha\phi_0 Q^2}{r_+((1+\alpha\phi_0^2)Q^2-(1+\alpha\phi_0^2)^2r_+^2)},\nonumber\\
&&\delta_1=-r_+\phi_1^2,\quad v_1=-\frac{e^{-\delta_0}Q}{r_+^2(1+\alpha\phi_0^2)}.
\end{eqnarray}
This approximate solution involves  two  parameters of  $\phi_0=\phi(r_+)$ and $\delta_0=\delta(r_+)$, which will be
found when  matching  (\ref{apsc-1})-(\ref{aps-4}) with the asymptotic  solutions in the far region
\begin{eqnarray}\label{ncoef-e}
m(r)&=&M-\frac{Q^2+Q_s^2}{2r}+\ldots,~\bar{\phi}(r)=\phi_\infty+\frac{Q_s}{r}+\ldots, \nonumber \\
\delta(r)&=&\frac{Q_s^2}{2r^2}+\ldots,~v(r)=\Phi+\frac{Q}{r}+\ldots, \label{insol}
\end{eqnarray}
where  $Q_s$ and $\Phi$ denote the scalar charge and the electrostatic potential, in addition to the ADM mass $M$ and the electric charge $Q$.
For simplicity, we choose  $\phi_\infty=0$.
\begin{figure*}[t!]
   \centering
   \includegraphics{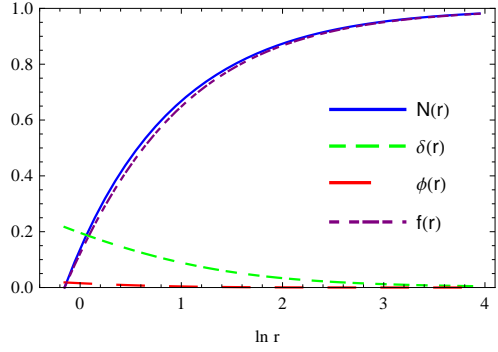}
\caption{A scalarized charged black hole solution  with $\alpha=8.083$ located in the $n=0 (\alpha\ge 8.019)$ fundamental branch in the EMS theory.
This is plotted as a function of $\ln r$ on and outside the horizon at $\ln r$= $\ln r_+=-0.154$. }
\end{figure*}
\begin{figure*}[t!]
  \includegraphics{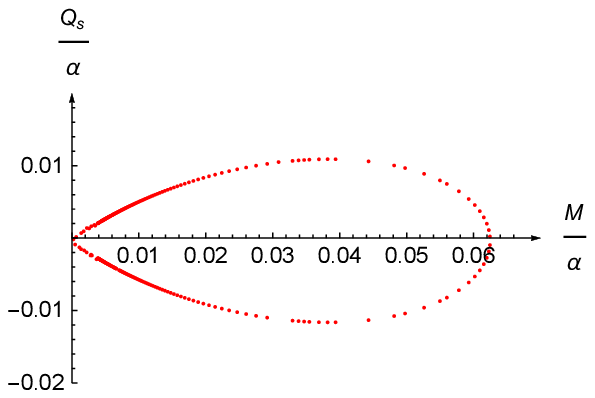}
  \hfill%
   \includegraphics{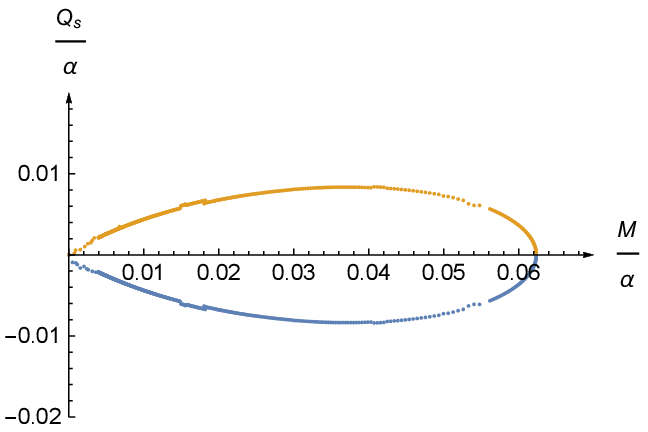}
\caption{Two similar graphs for  scalar charge $Q_s/\alpha$ vs $M/\alpha$  for the $n=0$ black hole. The $M/\alpha$-axis represents the unstable RN black hole with $q=0.7$ for $0<M/\alpha<0.06$  and  the stable RN black hole for  $M/\alpha>0.06$.
(Left) exponential coupling and (Right) quadratic coupling.}
\end{figure*}

Now, let us  display a numerical solution  with the coupling constant $\alpha = 8.083$  locating on the $n=0(\alpha \ge 8.019)$ fundamental branch in Fig. 2 by solving (\ref{neom1})-(\ref{neom4}) together with $q=0.7$ numerically. It is worth noting that the $n=1(\alpha\ge 40.84)$, $2(\alpha\ge 99.89)$
black holes  take the similar forms as the $n=0$ case. Actually, we need to obtain  hundreds of numerical solution depending $\alpha$ to compute quasinormal modes for full perturbations to each scalarized black hole.

On the other hand, we solve Eq.(\ref{neom1})
after replacing  $e^{\alpha (\bar{\phi})^2}$ with  $1+\alpha (\bar{\phi})^2$  and  Eq.(\ref{neom4}) after inserting $e^{\alpha(\bar{\phi})^2}$ at the first term to obtain the scalarized RN black holes in the EMS theory with exponential coupling.
From Fig. 3, we find that the fundamental branch ($n=0$) of exponential coupling is nearly the same as  that of quadratic coupling.
Here, both  the $n=0$ branches are  defined  from 0 to $\frac{M}{\alpha}=0.5/8.019\approx0.06$ where the RN black holes are unstable. For $M/\alpha>0.06$, the scalar hair (scalar charge $Q_s$) disappears
and the branch merges with the stable RN branch.

\section{EMS theory with  scalar mass term}
 Recently, it was shown that the introduction of a scalar mass term  has a significant influence on the bifurcation points
where the scalarized black holes branch out of the Schwarzschild black hole in the EGBS theory~\cite{Macedo:2019sem,Doneva:2019vuh}.
In this section, we wish to explore how  the introduction of  a specific mass term of $V_\phi=2m^2_\phi\phi^2$ in the EMS theory affects the bifurcation points
where the scalarized charged black holes branch out of the RN black hole with $q=0.418$.
In general, the presence of a massive scalar term affects significantly the stability of RN black hole and in turn the existence of scalarized charged black holes.
A choice of scalar mass $m^2_\phi=\alpha/\beta$ is quite interesting because it does not belong to an independent mass term, but it is given by the combination of coupling parameter $\alpha$ and mass parameter $\beta$. This choice would provide a compact result on the stability.

As a first step, we have to analysis the stability of RN black hole in the EMS theory with mass term based on the perturbed scalar equation
\begin{equation}
\Big(\bar{\square}-\frac{\alpha}{\beta}+ \frac{\alpha Q^2}{r^4}\Big) \delta \varphi= 0
\end{equation}
because two other linearized equations remain Einstein-Maxwell system for $\bar{\phi}=0$ case.
In this case,  a radial part of the scalar perturbed  equation takes the form
\begin{equation} \label{sch-2}
\frac{d^2u}{dr_*^2}+\Big[\omega^2-V(r)\Big]u(r)=0.
\end{equation}
Here the scalar potential $V(r)$ is given by
\begin{equation} \label{pot-c}
V(r)=\tilde{N}(r)\Big[\frac{2M}{r^3}+\frac{l(l+1)}{r^2}+\frac{\alpha}{\beta}-\frac{2Q^2}{r^4}-\alpha\frac{Q^2}{r^4}\Big].
\end{equation}
We focus on the $l=0$ mode only since the $s(l=0)$-mode is allowed for the scalar perturbation and it plays the important role in testing the stability of the RN black hole. Also, we emphasize  that $V(r) \to \alpha/\beta$ (positive) as $r\to \infty$, contrasting  to the massless case of  $V_{\rm ml}(r)\to 0$ in the EMS theory. This implies that we could not derive the sufficient condition for instability of $\int^\infty_{r_+} dr V_{\rm ml}(r)/\tilde{N}(r)<0$ in the EMS theory because of $\int^\infty_{r_+} dr V(r)/\tilde{N}(r)\to\infty$.

On the other hand,  observing the potential (\ref{pot-c}) carefully,  the positive definite potential without negative region (sufficient condition for stability)  could be implemented  by imposing the bound
\begin{equation}
\frac{V(r)}{\tilde{N}(r)} \ge 0 \to \beta \le g(r,\alpha)= \frac{\alpha r^4}{Q^2(\alpha+2)-2Mr},\label{c-po}
\end{equation}
which guarantees a stable RN black hole. This is so because $\tilde{N}(r)\le 0$ for $r\in[r_+,\infty]$.
\begin{figure*}[t!]
   \centering
  \includegraphics{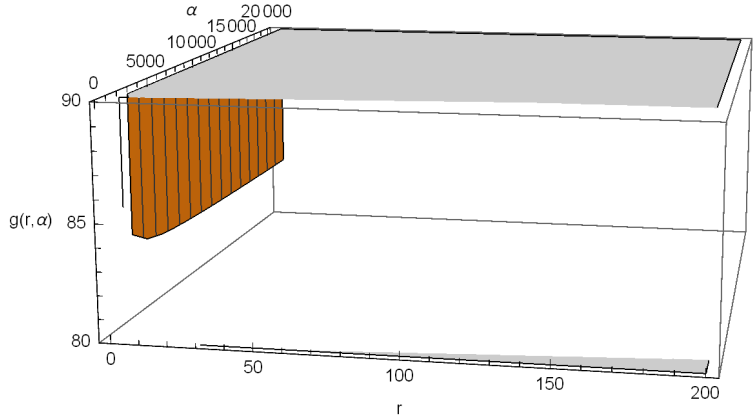}
\caption{A 3D graph of function $g(r,\alpha)$ for $r\in[r_+=1.995,200]$ and $\alpha\in[0.01,20,000]$. Its minimum stays  near $r=r_+=1.9943$ as $\alpha$ increases.
The gray strip along the $r$-axis indicates negative region of $g(r,\alpha)$  and so, it is excluded from consideration. }
\end{figure*}
In Fig. 4, we observe the behavior  of $g(r,\alpha)$ function. Minimum value of $g(r,\alpha)$ appears `83' around $r=r_+$ for $\alpha=20,000$.
Explicitly, the stability bound can be obtained from $g(r,\alpha)$ and Fig. 4 as
\begin{equation} \label{st-con}
\beta \le \frac{r_+^4}{Q^2}=83.217, {\rm as}~\alpha \to \infty,
\end{equation}
where any scalarized charged black holes could not be obtained  for any $\alpha$ because the appearance of the scalarized charged  black holes is closely related to the
instability of the RN black hole~\cite{Myung:2018vug}.

Unfortunately, it is hard to obtain the instability bound from the potential (\ref{pot-c}) directly.
First of all, we wish to find the negative region of potential outside the horizon because it may show a signal of instability.
Guided by the stability condition (\ref{st-con}), one expects that the negative region appears for $\beta>83.217$ and $\alpha<\infty$.
However, some potentials with negative region near the horizon  do not always  imply the instability. A truly criterion to determine whether a black hole
is stable or not  against the massive scalar perturbation depends on whether the time-evolution of the perturbation is decaying or not.
The linearized  equation (\ref{sch-2}) around a RN black hole may allow for a growing (unstable) mode like $e^{\Omega t}(\Omega>0)$ of the scalar perturbation and thus, it indicates the instability of the black hole.
Therefore, we solve (\ref{sch-2}) directly with appropriate boundary conditions.
From Fig. 5, we read off the thresholds  $\alpha_{\rm th}(\beta)$ of instability depending on $\beta$.
Importantly, the instability bound is determined  numerically by
\begin{equation}
\alpha(\beta) \ge \alpha_{\rm th}(\beta) \label{In-cond}
\end{equation}
where $ \alpha_{\rm th}(\beta)=$174.22(200), 132.51(240), 102.29(300), 82.76(380), and 68.41(500) is exactly the same as the  first bifurcation point $\alpha_{n=0}(\beta)$ which is determined when
solving the static perturbed equation (\ref{sch-2}) with $\omega=0$.
\begin{figure*}[t!]
   \centering
   \includegraphics{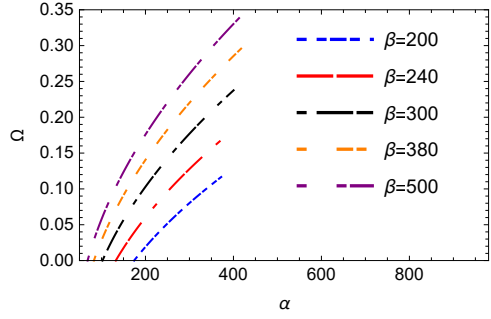}
\caption{Five graphs of $\Omega$ in $e^{\Omega t}$ vs $\alpha$ to determine the thresholds of instability [$\alpha_{\rm th}(\beta)$] which are the crossing points at $\alpha$-axis. We read off those as $\alpha_{\rm th}(\beta)$=174.22(200), 132.51(240), 102.29(300), 82.76(380), and 68.41(500).}
\end{figure*}
We find that $\alpha_{n=0}(\beta)$ decreases as $\beta$ increases. It is conjectured  that $\alpha_{n=0}(\beta)\to \infty$, as $\beta \to 83.217$.
In other words, we show that there is no unstable RN black holes for the case of $\beta\le 83.217$, where
any scalarized charged black holes could not be found for any $\alpha$.

Importantly, it is noted that the RN black hole  is allowed for any value of $\alpha$,
whereas a scalarized charged  black hole solution
may exist only for $\alpha(\beta)\ge \alpha_{\rm th}(\beta)$ for $\beta > 83.217$.
A close connection always exists between the instability of a RN black hole and appearance of  the $n=0$  scalarized charged  black hole in the EMS theory with massive scalar term.

Now, let us  derive  the $n=0$  scalarized charged black hole which corresponds  to the  $q=0.418$ and $\alpha(\beta=200) \ge 174.22$ case.
Adopting the metric ansatz (\ref{nansatz}), Eqs.(\ref{neom1})-(\ref{neom4}) get modified to  include a scalar mass  term.
The approximate solution in the near horizon is the same form as in (\ref{apsc-1})-(\ref{aps-4})
with the same coefficients as $\delta_1$ and $v_1$ in (\ref{ncoef}) and two different coefficients
\begin{eqnarray}
&&m_1=\frac{Q^2}{2r_+^2(1+\alpha\phi_0^2)}+\frac{\alpha r_+^2\phi_0^2}{2\beta},\label{ncoef1}\\
&&\phi_1=\frac{\alpha\phi_0 Q^2-\alpha(1+\alpha\phi_0^2)\phi_0r_+^4/\beta }{r_+(1+\alpha\phi_0^2)\Big[Q^2-r_+^2(1+\alpha\phi_0^2)(1-\alpha\phi_0^2r_+^2/\beta)\Big]},\label{ncoef2}
\end{eqnarray}
which lead to  (\ref{ncoef}) in the massless limit of $\beta \to \infty$.

On the other hand, the asymptotic solution in the far region takes the different form
\begin{eqnarray}
m(r)&=&M-\frac{Q^2}{2r}-\frac{Q_s^2e^{-2\sqrt{\frac{\alpha}{\beta}}r}}{2r^{1+2M\sqrt{\frac{\alpha}{\beta}}}}+\ldots,\nonumber\\
v(r)&=&\Phi+\frac{Q}{r}+\frac{e^{-2\sqrt{\frac{\alpha}{\beta}}r}}{r^{
2M\sqrt{\frac{\alpha}{\beta}}}}\frac{QQ_s^2}{2\sqrt{\alpha\beta}r^4}+\ldots,\nonumber\\
\delta(r)&=&Q_s^2(2\sqrt{\alpha/\beta})^{2+2M\sqrt{\alpha/\beta}}\Gamma[-2
-2M\sqrt{\alpha/\beta},2\sqrt{\alpha/\beta}r]+\ldots, \nonumber \\
\phi(r)&=&\frac{Q_se^{-\sqrt{\alpha/\beta}r}}{r^{1+M\sqrt{\alpha/\beta}}}+\ldots\label{insol},
\end{eqnarray}
which lead to  (\ref{ncoef-e}) except $\bar{\phi}(r)$ in the massless limit of $\beta \to \infty$. This means that all asymptotic forms  are changed under the inclusion of scalar mass term.
We wish to display a numerical solution with $\alpha=198.34$ belonging to the $n=0$ fundamental branch   in Fig. 6 by solving (\ref{neom1})-(\ref{neom4}) together with mass term. Here we observe that $N(r)$ and $\delta(r)$ are similar to those in Fig. 2 of the EMS theory, while $\bar{\phi}(r)$ shows a different asymptotic behavior from the scalar in the EMS theory. 
\begin{figure*}[t!]
   \centering
   \includegraphics{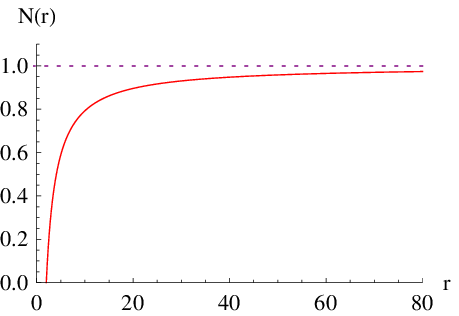}
   \hfill%
   \includegraphics{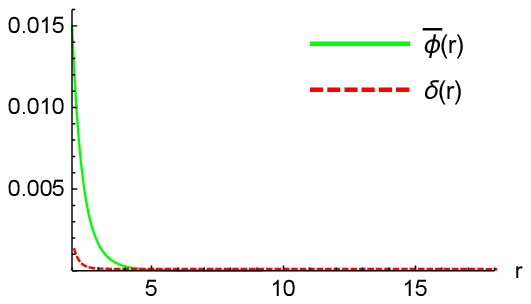}
\caption{Plots of a scalarized charged black hole with $\alpha=198.34$ for the $n=0$ fundamental branch of $\alpha(\beta=200) \ge 174.22$ and $q=0.418$  in the EMS theory with massive scalar. These are plotted as a function of $r$ on and outside the horizon at $r_+=1.9943$. }
\end{figure*}

\section{Full linearized theory}

We consider  the full perturbed fields around the background quantities
\begin{eqnarray}
  g_{\mu\nu}= \bar{g}_{\mu\nu}+h_{\mu\nu},~
  A_\mu =\bar{ A}_\mu(r)+a_{\mu},~
  \phi= \bar{\phi}(r) +\delta \phi. \label{per-3}
\end{eqnarray}
Plugging (\ref{per-3}) into Eqs.(\ref{equa1})-(\ref{s-equa}) leads to complicated  linearized equations.
Considering  ten degrees of freedom for $h_{\mu\nu}$, four for $a_{\mu}$, and one for $\delta \phi$ initially,
the EMS theory describing a massless scalar and massless vector-tensor propagations provides  five (1+2+2=5) physically propagating modes on the scalarized black hole background.
The stability analysis should be based on these physically propagating fields as the solutions to the linearized equations.
In a spherically
symmetric background (\ref{nansatz}), the perturbations can be decomposed into spherical harmonics $Y_{lm}(\theta,\varphi)$ with multipole index
$l$ and azimuthal number $m$. This decomposition
splits  the tensor-vector perturbations into ``axial (A) part" and ``polar (P) part".

We expand the metric perturbations in tensor spherical
harmonics under the  Regge-Wheeler gauge, providing six degrees of freedom.  The axial part $h^{\rm A}_{\mu\nu}(t,r,\theta,\varphi)$ is composed of two radial modes $h_0(r)$ and $h_1(r)$ and
 the polar part $h^{\rm P}_{\mu\nu}(t,r,\theta,\varphi)$ takes  four radial modes [$H_0(r),H_1(r),H_2(r),K(r)$] with time-dependence $e^{-i\omega t}$.
Similarly, we decompose  the vector perturbations into the axial vector $a^{\rm A}_{\mu}(t,r,\theta,\varphi)$ with single  mode  $u_4(r)$ and the polar vector  $a^{\rm P}_{\mu}(t,r,\theta,\varphi)$ with two modes $u_1(r)$ and $u_2(r)$, giving three degrees of freedom.
Lastly, we have a polar scalar perturbation as
\begin{eqnarray}
\delta\phi(t,r,\theta,\varphi)=\int d\omega e^{-i\omega t}\sum_{l,m}\delta \phi_1(r)Y_{lm}(\theta,\varphi).
\end{eqnarray}
We note that  the linearized equations could be split into axial  and  polar parts.

In general, the axial part is composed of two coupled equations for  Maxwell $\hat{F}(u_4)$ and Regge-Wheeler $\hat{K}(h_0,h_1)$,
\begin{eqnarray} \label{axial-1}
 \Big[\frac{d^2}{dr_{*}^2}+\omega^2\Big]\hat{F}(r)&=&V^{\rm A}_{\rm FF}(r)\hat{F}(r)+V^{\rm A}_{\rm FK}(r)\hat{K}(r), \\
 \Big[\frac{d^2}{dr_{*}^2}+\omega^2\Big]\hat{K}(r)&=&V^{\rm A}_{\rm KK}(r)\hat{K}(r)+V^{\rm A}_{\rm KF}(r)\hat{F}(r),\label{axial-2}
\end{eqnarray}
where the potentials are given by
\begin{eqnarray}
  V^{\rm A}_{\rm FF}(r) &=& \frac{N}{r^2 e^{2\delta}}\Big[e^{2\delta}r^2\left(4(1+\alpha\bar{\phi}^2)-\frac{\alpha^2\bar{\phi}^2}{1+\alpha\bar{\phi}^2}\right)(v')^2+l(l+1)
  +\frac{\alpha rN\bar{\phi}'}{1+\alpha\bar{\phi}^2}\left(\frac{r\bar{\phi}'}{1+\alpha\bar{\phi}^2}-2\bar{\phi}\right)\Big], \nonumber\\
  V^{\rm A}_{\rm FK}(r) &=& V^{\rm A}_{\rm KF}(r)=-\frac{2\sqrt{1+\alpha\bar{\phi}^2}e^{-\delta}(l-1)(l+2)N v'}{r}, \nonumber\\
  V^{\rm A}_{\rm KK}(r) &=& \frac{N}{r^2 e^{2\delta}}\Big[(l-1)(l+2)-r N'+N(2+r\delta)\Big].
\end{eqnarray}
Here the tortoise coordinate $r_*\in(-\infty,\infty)$ is defined by the relation of $dr_*/dr= e^\delta/N$.
At this stage, it is worth noting that in the limits of $\bar{\phi}=\delta=0$, $V^{\rm A}_{\rm FF}(r)$, $V^{\rm A}_{\rm FK}(r)$, and $ V^{\rm A}_{\rm KK}(r)$ recovers those for the RN black hole in the Einstein-Maxwell theory~\cite{Chandrasekhar:1979iz}. Here, we will derive the quasinormal modes propagating around $ n=0,1,2$ scalarized  black holes by solving the two coupled equations directly.

On the other hand, the polar part is composed of  six coupled equations for Zerilli (3), Maxwell (2),  and scalar (1)  as
\begin{eqnarray}
 K'(r)&=&-\left(\frac{l(l+1)+2N+2r N'-2}{2r^2}+e^{2\delta}(1+\alpha\bar{\phi}^2)v'^2+N\bar{\phi}'(r)^2\right)H_1(r)\nonumber\\
 &&+\frac{H_0(r)}{r}+\left(\frac{N'}{2N}-\frac{1}{r}-\delta'\right)K(r)-\frac{2\bar{\phi}'}{r}\delta \phi_1(r),\label{pol-eq1} \\
 H_1'(r)&=&-\frac{4i(1+\alpha\bar{\phi}^2)v'}{\omega}f_{12}(r)-\frac{H_0(r)+K(r)}{N}+\left(\delta'-\frac{N'}{N}\right)H_1(r), \label{pol-eq2}\\
H_0'(r)&=&\left(\frac{1}{r}+2\delta'-\frac{N'}{N}\right)\Big[H_0(r)-K(r)\Big]
+\frac{4e^{2\delta}(1+\alpha\bar{\phi}^2)v'}{N}f_{02}(r)+\frac{2\bar{\phi}'}{r}\delta\phi_1(r) \label{pol-eq3}\\
&&+\left(\frac{e^{2\delta}\omega^2}{N}-e^{2\delta}(1+\alpha\bar{\phi}^2)v'^2-N\bar{\phi}'^2-\frac{l(l+1)}{2r^2}-\frac{N+rN'-1}{r^2}\right)H_1(r),\nonumber\\
f_{02}'(r)&=&v'K(r)+\frac{2\alpha\bar{\phi}V'}{r(1+\alpha\bar{\phi}^2)}\delta\phi_1(r)+\left(\frac{l(l+1)ie^{-2\delta}N}{r^2\omega}-i\omega\right)f_{12}(r),\label{pol-eq4}\\
f_{12}'(r)&=&-\frac{i\omega e^{2\delta}}{N^2}f_{02}(r)+\left(\delta'-\frac{2\alpha\bar{\phi}\bar{\phi}'}{1+\alpha\bar{\phi}^2}
-\frac{N'}{N}\right)f_{12}(r),\label{pol-eq5}\\
\delta\phi''_1(r)&=&\Big[\frac{e^{2\delta}\alpha(-1+3\alpha\bar{\phi}^2)v'^2}{N(1+\alpha\bar{\phi}^2)}+\frac{l(l+1)}{r^2N}-\frac{e^{2\delta}\omega^2}{N^2}
+\frac{N'+N(-\delta'+4r\bar{\phi}'^2)}{rN}\Big]\delta\phi_1(r)\nonumber\\
&&+\left(\delta'-\frac{N'}{N}\right)\delta\phi'_1(r)+\frac{2il(l+1)\alpha\bar{\phi}v'}{r\omega}f_{12}(r)
+\frac{4e^{2\delta}(1+\alpha\bar{\phi}^2)r v'\bar{\phi}}{N}f_{02}(r)\label{pol-eq6}\\
&&-\frac{r\left(e^{2\delta}\alpha\bar{\phi}v'^2+(N'-2N\delta')\bar{\phi}'\right)}{N}H_0(r)
+\frac{2re^{2\delta}\alpha\bar{\phi}v'^2}{N}K(r)\nonumber
\end{eqnarray}
with $H_2(r)=H_0(r)$ and $f_{01}(r)=i\omega f_{12}(r)+f'_{02}(r)$.
Interestingly, these coupled equations describe three physically propagating modes.

\section{Stability Analysis of $n=0,1,2$ black holes}

 First of all,  we wish to mention briefly  why the $n=0$  black hole is stable (unstable) against radial  perturbations for the exponential (quadratic) coupling in the EGBS theory by providing two kinds of  potentials. It is well known that the radial perturbations for $l=0,1$-modes are equivalent to the full perturbations for the same modes. However, this is not true for higher modes of $l=2,3,4,\cdots$ which are necessary to introduce the full perturbations.
The EGBS theory~\cite{Doneva:2017bvd} is given  by
\begin{equation}
S_{\rm EGBS}=\frac{1}{16 \pi}\int d^4 x\sqrt{-g}\Big[ R-2\partial_\mu \phi \partial^\mu \phi
+\lambda^2f(\phi){\cal R}^2_{\rm GB}\Big],\label{Action-GB}
\end{equation}
where $\lambda$ is the Gauss-Bonnet coupling constant and $f(\phi)$ is the coupling function defined as
\begin{eqnarray}
{\rm exponential}: f(\phi)=\frac{1}{12}(1-e^{-6\phi^2}); \quad {\rm quadratic}: f(\phi)=\frac{1}{2}\phi^2.
\end{eqnarray}
When solving two linearized scalar equations with static ansatz,
one   obtains a discrete spectrum of parameter $\lambda$ as $M/\lambda= \{0.587, 0.226, 0.140\ldots\}$,
which describes the $n=0,1,2,\cdots$ scalarized black holes~\cite{Myung:2018iyq}.
From Fig. 7, we observe  that the fundamental branch  of $n=0$ black hole is a finite region of  $0<M/\lambda<0.587$ in the exponential coupling,
while it is just a band with bandwidth of $0.587<M/\lambda<0.636$  for quadratic coupling~\cite{Silva:2017uqg,Blazquez-Salcedo:2018jnn}. It is important to note that the latter locates   on the stable Schwarzschild black hole bound (outside the fundamental branch for exponential coupling).  This points out  one of differences between exponential and quadratic couplings in the EGBS theory.
\begin{figure*}[t!]
  \includegraphics{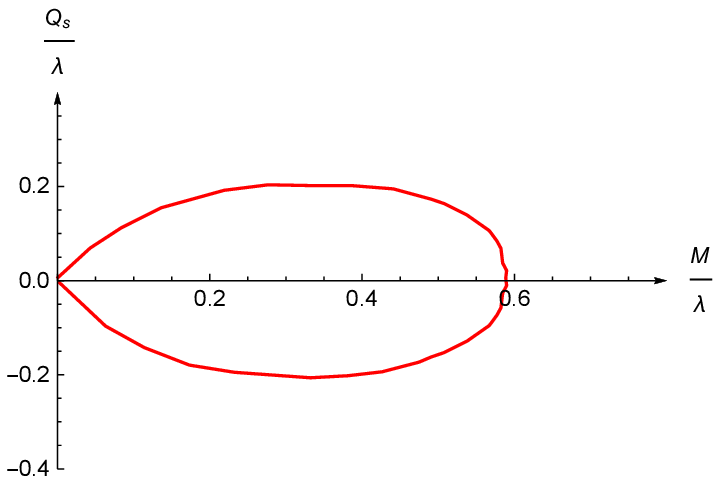}
  \hfill%
   \includegraphics{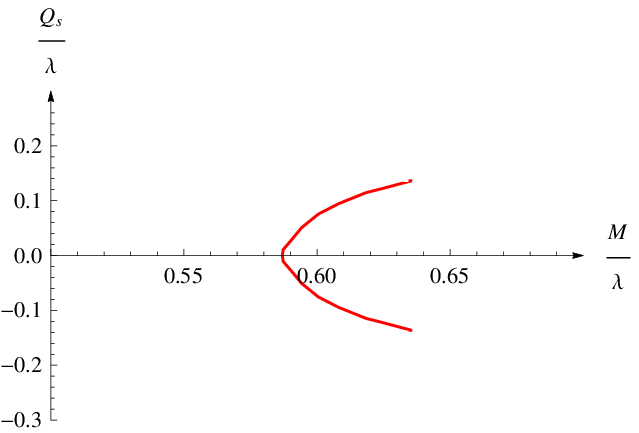}
\caption{Two different graphs of scalar charge $Q_s/\lambda$ vs $M/\lambda$ for the $n=0$ black hole with $r_+=1.174$.
 The $M/\lambda$-axis represents the unstable Schwarzschild  black hole for $0<M/\lambda<0.587$ and
and the stable Schwarzschild black hole  for  $M/\lambda>0.587$~\cite{Myung:2018iyq}.
(Left) exponential coupling has a finite region of $0<M/\lambda<0.587$ which is the unstable bound for Schwarzschild black hole,
 while (Right) quadratic coupling has a band with bandwidth of 0.587$<M/\lambda<$0.636 which exits within the stable bound for Schwarzschild black hole.}
\end{figure*}
Introducing radial (spherically symmetric) perturbations around the scalarized black holes as
\begin{eqnarray}
ds^2_{\rm EGBS}&=&-N(r)e^{-2\delta(r)}\left(1+\epsilon H_0\right)dt^2+\frac{dr^2}{N(r)\left(1
+\epsilon H_1\right)}+r^2(d\theta^2+\sin^2\theta d\varphi^2),\nonumber\\
\phi&=&\bar{\phi}(r)+\epsilon\delta\phi, \label{ESGB-metric}
\end{eqnarray}
a decoupling process makes  a single second order  equation for scalar perturbation~\cite{Blazquez-Salcedo:2018jnn,Minamitsuji:2018xde}
\begin{eqnarray}
g(r)^2\frac{\partial^2\delta\phi}{\partial t^2}-\frac{\partial^2\delta\phi}{\partial r^2}
+C_1\frac{\partial\delta\phi}{\partial r}+U(r)\delta\phi=0,\label{ESGB-pert1}
\end{eqnarray}
where  $g(r)$, $C_1(r)$ and $U(r)$ are functions of  $N(r)$, $\delta(r)$ and $\bar{\phi}(r)$~\cite{Blazquez-Salcedo:2018jnn}.
Considering a further separation of perturbed scalar $\delta\phi(t,r)=\delta\phi(r)e^{-i\omega t}$,
we obtain the Schr\"{o}dinger equation for scalar perturbation
\begin{eqnarray}
\frac{d^2Z}{d r_*^2}=\Big[V(r)-\omega^2\Big]Z,\label{ESGB-pert2}
\end{eqnarray}
where $r_*$ is the tortoise coordinate and  a redefined scalar perturbation  $Z(r)$ reads as
\begin{eqnarray}
r_*=\int_{r_+}^\infty g(r) dr,\quad Z(r)=\frac{\delta\phi(r)}{C_0(r)}.\label{ESGB-Z}
\end{eqnarray}
\begin{figure*}[t!]\label{potv1}
\includegraphics{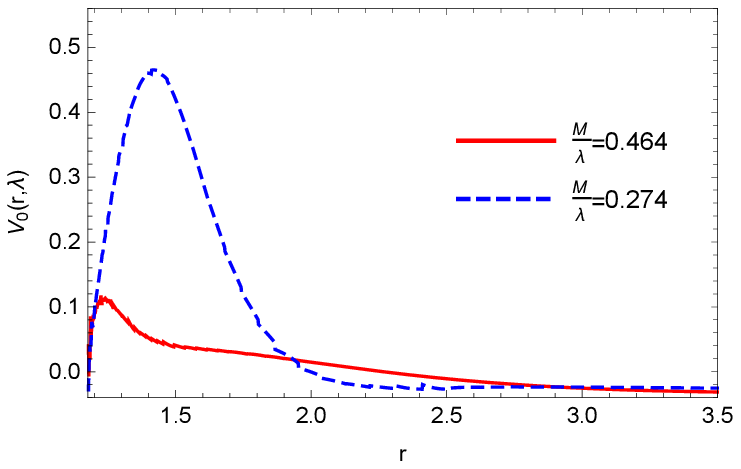}
\hfill%
\includegraphics{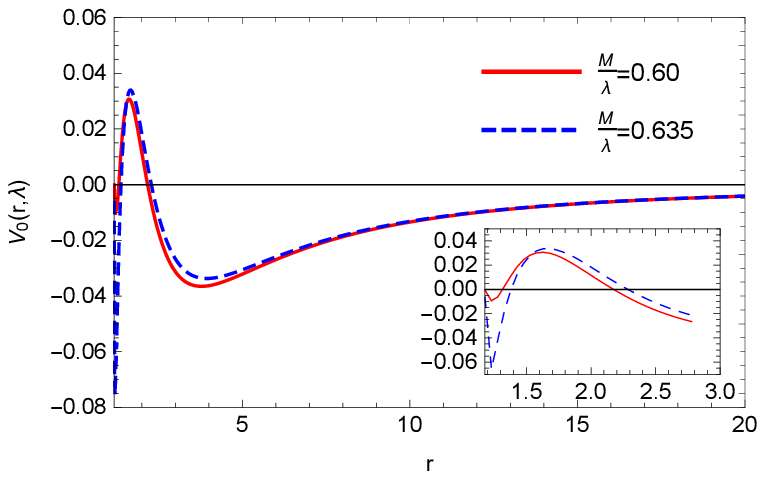}
\caption{Two scalar potential graphs of  $V_0(r,\lambda)$ for $s$-mode scalar around the $n=0$ black hole with
 horizon radius $r_+=1.174$. (Left) exponential coupling. (Right) quadratic coupling. The magnification of the enclosed region shows the specific potential behaviors   just outside the horizon, indicating   negative-positive-negative regions.
}
\end{figure*}
Importantly, the potential is given by
\begin{eqnarray}
V(r)&=&\frac{U(r)-C_1'(r)}{g(r)^2}+\frac{C_1g'(r)+g''(r)}{g(r)^3}-\frac{2g'(r)^2}{g(r)^4}.\label{ESGB-V}
\end{eqnarray}
In Fig.~8, we plot  the potentials $V_0(r,\lambda)$ for $l=0$-scalar mode around the $n=0$ black hole in the EGBS theory with exponential and quadratic couplings.
It is obvious that the potential for exponential coupling is positive outside the horizon,
while the potential for quadratic coupling develops negative-positive-negative regions outside the horizon, leading to $\int_{r_+}^\infty V_0(r)g(r)dr<0$ [sufficient condition for instability]. This is the other of differences between exponential and quadratic couplings. Thus, the endpoint of unstable Schwarzschild black holes may be  the stable $n=0$ black hole in the EGBS theory with exponential coupling only.

Now let us turn to the stability analysis for the $n=0,1,2$ black holes in the EMS theory.
The stability analysis may  be performed by getting quasinormal frequency of $\omega=\omega_{r}+i\omega_i$ in $e^{-i\omega t}$ when  solving the linearized equations with appropriate boundary conditions at the outer horizon: ingoing waves and at infinity: purely outgoing waves.  We will compute the lowest quasinormal modes of the scalarized black holes by making use of a
 reasonable  numerical background and the linearized equations (\ref{axial-1})-(\ref{axial-2}) for axial part and the linearized equations (\ref{pol-eq1})-(\ref{pol-eq6}) for polar part.
To compute the quasinormal modes, we use a direct-integration method~\cite{Blazquez-Salcedo:2016enn}.

Usually, a positive definite potential $V(r)$ without any negative region guarantees the stability of black hole.
On the other hand, a sufficient condition for instability is given by $\int^{\infty}_{r_+} dr [e^\delta V(r)/N(r)] <0$~\cite{Dotti:2004sh} in accordance with the existence of  unstable modes.
However, some potentials with negative region  near the outer horizon whose integral is positive ($\int^{\infty}_{r_+} dr [e^\delta V(r)/N(r)] >0$) may  not imply a definite instability.  To determine the instability of the $n=0,1,2$ black holes clearly,  one has to solve all linearized equations for physical perturbations numerically.
Accordingly, the criterion to determine whether a black hole is stable or not against the physical perturbations is whether the time evolution  $e^{-i\omega t}$  of the perturbation is decaying or not. If $\omega_i<0(>0)$, the black hole is stable (unstable), irrespective of any value of $\omega_r$.
However, it is not an easy task to carry out the stability of scalarized charged black holes because these black holes comes out as  not an analytic solution but   numerical solutions. To have a reasonable numerical background, it needs to obtain hundreds  of numerical solutions in the each branch.
It is convenient to  classify the linearized equations according to multiple index of $l=0,1,2,\cdots$ because  $l$ determines number of physical fields at the axial and polar sectors.

\subsection{$l=0$ case: $n=0,1,2$ black holes}

For $l=0$($s$-mode), the linearized equation obtained from the polar part is given entirely  by a scalar equation ($S^{\rm P}_0=r \delta \phi_1$)
\begin{eqnarray}
\Big[\frac{d^2}{dr_{*}^2}+\omega^2\Big]S^{\rm P}_0-V^{\rm P}_{0}(r,\alpha)S^{\rm P}_0=0,
\end{eqnarray}
where the potential $V^{\rm P}_{\rm 0}(r,\alpha)$ is given by~\cite{Fernandes:2019rez}
\begin{eqnarray}
V^{\rm P}_{\rm 0}(r,\alpha)&=&\frac{N}{e^{2\delta}r^2}\Big[(N+r(N'-N\delta')-1)\left(2r^2\bar{\phi}'^2-\frac{4\alpha\bar{\phi}\bar{\phi}'r^2}{1+\alpha\bar{\phi}^2}
-\frac{1+\alpha+\alpha\bar{\phi}^2(2-3\alpha+\alpha\bar{\phi}^2)}{(1+\alpha\bar{\phi}^2)^2}\right)\nonumber\\
&&+1-N-2r^2(\bar{\phi}')^2\Big],
\label{potl=0}
\end{eqnarray}
which is the same form as that obtained by taking radial perturbations~\cite{Fernandes:2019rez}.
We display three scalar potentials $V^{\rm P}_{\rm 0}(r,\alpha)$  in Fig. 9 for $l=0$ case around the $n=0$ black hole. The whole potentials are positive definite except that the $\alpha=8.65$ case has  negative region near the horizon. It does not  represent instability  because this is near the threshold of instability. Actually,  the $n=0$ black hole is stable against the $l=0$ scalar perturbation.
 We confirm it  from Fig. 10 that  the imaginary frequency is negative  for $\alpha \ge 8.019$, implying a stable $n=0$ black hole.
 This means that the endpoint of unstable RN black holes with $\alpha>8.019$
is the  $n=0(\alpha \ge 8.019)$ scalarized charged  black hole with the same $q$.
This is one of our main results.

Now let us turn to the stability issue of the $n=1,2$  black holes.
We observe from Fig. 11 that  $\int^{\infty}_{r_+} dr [e^\delta V(r)/N(r)] <0$ for the $n=1$ black hole, while the whole potentials are negative definite for the $n=2$ black hole.
This implies that the $n=1,2$ black holes are unstable against the $l=0$ scalar perturbation.
Clearly, the instability could be found  from Fig. 10 because their imaginary frequencies are positive.
Here the red curve denotes the unstable RN black holes as a function of $\alpha>8.091$.
Hereafter, we will perform  the stability analysis for higher multipoles on the $n=0$ black hole only because the $n=1,$ 2 black holes turned out to be unstable against the $l=0$ scalar perturbation. In other words, it seems meaningless  to carry out a further stability analysis for the unstable $n=1,$ 2 black holes.
\begin{figure}[htb]
   \centering
   \includegraphics{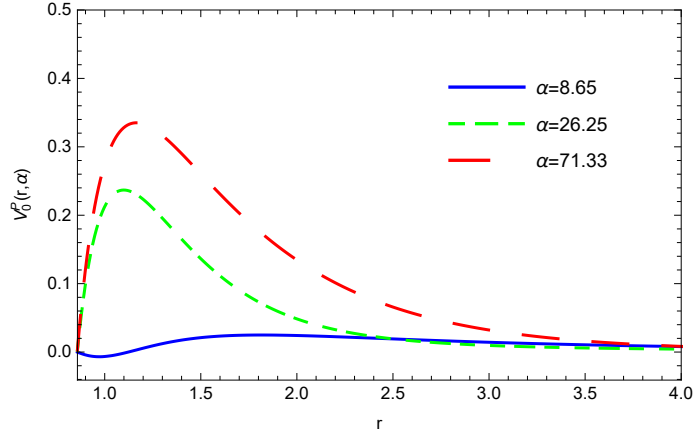}
\caption{Three scalar potential graphs $V^{\rm P}_0(r,\alpha)$  for $l=0$ mode around the $n=0(\alpha\ge 8.019)$ black hole.
The whole potentials are positive definite except that the $\alpha=8.65$ case having negative region near the horizon does not imply instability.}
\end{figure}
\begin{figure*}[t!]
   \centering
   \includegraphics{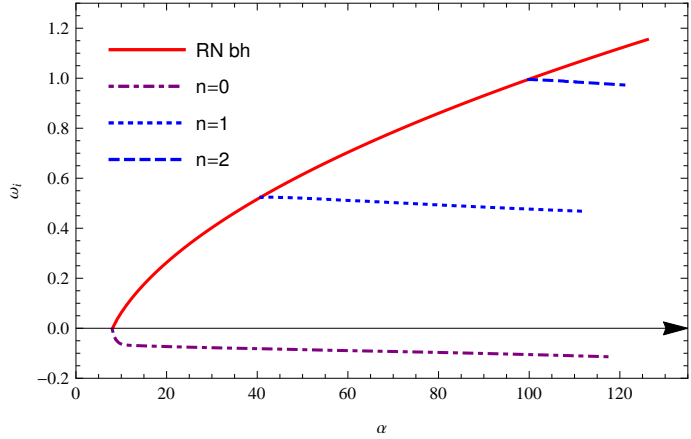}
\caption{ The negative imaginary frequency $\omega_i$ ($\omega_r=0$) as function of $\alpha$ appears for  the $l=0$ scalar  around the $n=0$ black hole,  implying the stability.
Two positive imaginary frequencies $\omega_i$ ($\omega_r=0$) are as functions of $\alpha$ for  the $l=0$ scalar around
the $n=1,2$ black holes, indicating the instability. A red solid curve with $q=0.7$ represents the quasinormal frequency of $l=0$ scalar mode as function of $\alpha$ around the RN black hole~\cite{Myung:2018vug}, showing the unstable  RN black holes for $\alpha>8.019$. }
\end{figure*}
\begin{figure*}[t!]
  \centering
   \includegraphics{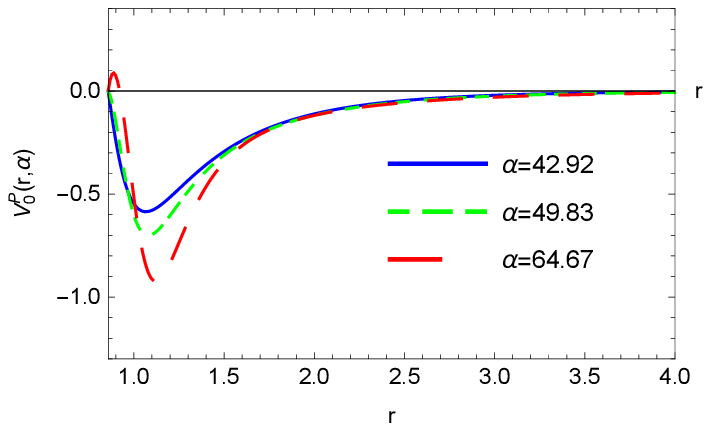}
   \hfill%
  \includegraphics{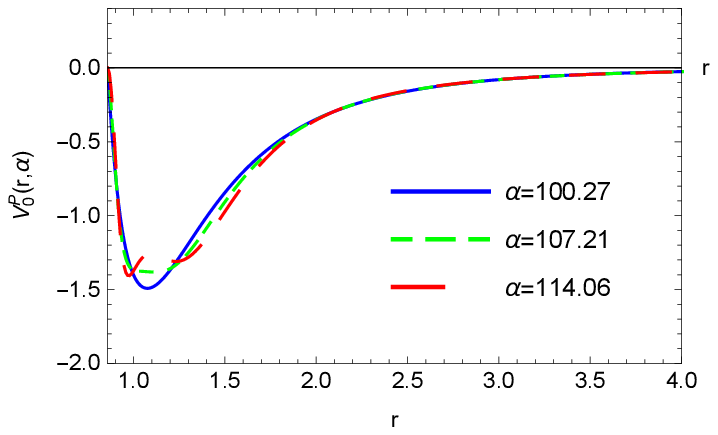}
\caption{Three scalar potential graphs $V^{\rm P}_0(r,\alpha)$ for  $l=0$ scalar around  (Left) $n=1(\alpha\ge 40.84)$ black hole and  (Right) $n=2 (\alpha\ge 99.89)$ black hole. }
\end{figure*}

\subsection{$l=1$ case: $n=0$ black hole}

In this case, we have three physical modes propagating around the $n=0$ black hole.
For $l=1$ case, the axial linearized equation around the $n=0$ black hole  is given by
\begin{equation}
\Big[\frac{d^2}{dr_{*}^2}+\omega^2\Big]Z^{\rm A}_1-V^{\rm A}_1(r,\alpha)Z^{\rm A}_1=0,
\end{equation}
where the potential takes the form
\begin{eqnarray}
V^{\rm A}_1(r,\alpha)=-\frac{e^{-2\delta} N}{r^2}\Big[&N&\Big(4-\alpha^2\bar{\phi}^2+\alpha r(\bar{\phi}^2)'-r^2(\alpha-4+2\alpha^2\phi^2)(\bar{\phi}')^2\Big)\nonumber \\
&-&6+4r N'+\alpha^2\bar{\phi}^2(1-r N')\Big]\label{potl=1a}
\end{eqnarray}
We find  that all potentials are  positive definite for the $n=0(\alpha\ge 8.019)$ black hole.
 This means that the $n=0$ black hole is stable against the axial $l=1$ vector perturbation.
 We confirm it by showing  that  $\omega_i$ is  negative, indicating a stable black hole.

Finally, we obtain the vector-led and scalar-led modes propagating around the $n$=0 black hole by solving  the polar $l=1$  linearized equations (\ref{pol-eq1})-(\ref{pol-eq6}).
We find that $\omega_i$ of vector-led mode around the $n=$0 is negative, implying a stable black hole.
Also, it is found that  $\omega_i$ of scalar-led mode  around the $n=0$ black hole is negative, implying a stable black hole.

\subsection{$l=2$ case: $n=0$ black hole}

First of all, we consider the axial part because of its simplicity.
The axial $l=2$ linearized equations are given by   two coupled equations for Regge-Wheeler-Maxwell system as shown in (\ref{axial-1})-(\ref{axial-2}).
Solving these coupled equation with boundary conditions leads to  negative quasinormal frequencies $\omega_i$ for $l=2$ vector-led mode  around  the $n=0$ black hole, implying stable black hole.
Also, we find  that the $n=0$ black hole is stable against the $l=2$ gravitational-led mode.

Finally, the polar  $l=2$ linearized equations are given by Eqs.(\ref{pol-eq1})-(\ref{pol-eq6}) with $l=2$.
Here we have three modes: vector-led, gravitational-led, and scalar-led modes.
We find  that all $\omega_i$ of these modes are negative, implying the stable $n=0$ black hole.

\section{Summary and Discussions}

First of all, it was shown in the EGBS theory that the $n=0$  black hole is stable against radial perturbations for the exponential coupling, while it  is unstable for the quadratic coupling. In the former case, the $n=0$ black hole could be regarded as  the endpoint of the evolution of unstable Schwarzschild black hole, whereas this is not the case for the latter.
We wish to point out  the differences between exponential and quadratic couplings for the $n=0$ black hole (fundamental blanch) in the EGBS theory.
We observe from Fig. 7 that  the fundamental branch  of $n=0$ black hole is a finite region of  $0<M/\lambda<0.587$ in the exponential coupling,
while it is just a band with bandwidth of $0.587<M/\lambda<0.636$  for quadratic coupling where locates  within the stable Schwarzschild black hole bound (beyond the fundamental branch for exponential coupling).  This is one difference between exponential and quadratic couplings in the EGBS theory.
Also, it is shown from Fig. 8 that the potential for exponential coupling is positive outside the horizon,
while the potential for quadratic coupling develops negative-positive-negative regions outside the horizon, leading to $\int_{r_+}^\infty V_0(r)g(r)dr<0$ [sufficient condition for instability]. This corresponds to the other difference between exponential and quadratic couplings for the $n=0$ black hole in the EGBS theory.

 Concerning the EMS theory with scalar mass $m^2_\phi=\alpha/\beta$, there is no unstable RN solution  with $q=0.418$ for $\beta \le 83.217$ as $\alpha \to \infty$.
This implies that for $\beta \le 83.217$, any scalarized charged black holes could not found for any $\alpha$. On the other hand,
we may develop the $n=0,1,2,\cdots$ scalarized charged black holes for the case of  $\beta > 83.217$ without limitation on number of  bifurcation points
even though the mass term changes  significantly the location of bifurcation points. We have found the $n=0$ scalarized charged black hole solution  from the EMS theory with scalar mass term whose metric functions are similar to those in the $n=0$ scalarized black hole obtained from the EMS theory. It is emphasized that the  scalar is different from that found in the EMS theory. However, the stability analysis of the $n=0$ scalarized charged black hole in the EMS theory with scalar mass term seems to be a difficult and complicated task and thus, we could not report its result on this work.
This is mainly due to difficulty in handling the asymptotic boundary conditions.

We have shown that  the $n=1(\alpha\ge 40.84),2(\alpha\ge 99.89)$ excited black holes are unstable against  against the $l=0$ scalar perturbation, while the $n=0(\alpha \ge 8.019)$ fundamental black hole is stable against all scalar-vector-tensor perturbations in the EMS theory with quadratic coupling.
In the latter, we found all negative  quasinormal frequencies ($\omega_i<0$) of $9=1(l=0)+3(l=1)+5(l=2)$ physical modes around the $n=0$ black hole.
In other words, we could not find any unstable modes from the $l=0,1,2$ scalar-vector-tensor  perturbations around the $n=0$ black hole.
Even though we have carried out the stability analysis on the $n=0$, 1, 2 black holes, we expect   from Fig. 10 that the other  higher excited ($n=$3, 4, 5,$\cdots$) black holes are unstable against the $s(l=0)$-mode scalar perturbation because their frequencies may  exist as further branches along the unstable RN black holes.
This is consistent with those for the EMS theory with exponential coupling~\cite{Myung:2018jvi}, but it contrasts to the $n=0$ scalarized black hole  found in the ESGB theory with quadratic coupling  when  making use of radial  perturbations~\cite{Blazquez-Salcedo:2018jnn}. Actually,  the $n=0$ black hole  found in the ESGB theory with exponential  coupling has a similar property found  in the EMS theory with exponential and quadratic couplings (See Figs. 3,~7).
This implies that the endpoint of unstable RN black holes with $\alpha>8.019$ is the $n=0$ scalarized black hole with the same $q=0.7$ in the EMS theory with quadratic and exponential couplings.

 \vspace{1cm}

{\bf Acknowledgments}

 This work was supported by the National Research Foundation of Korea (NRF) grant funded by the Korea government (MOE)
 (No. NRF-2017R1A2B4002057).
 
\end{document}